\documentstyle[floats,
pre,
epsfig,
aps,
]{revtex}
\parskip=\medskipamount
\preprint{Draft}
\begin{document}
\draft
%%%%%%%%%%%%%%%%%%%%%%%
\renewcommand{\textfraction}{0.0}
\renewcommand{\dblfloatpagefraction}{0.8}
\renewcommand{\topfraction}{1.0}
\renewcommand{\bottomfraction}{1.0}
\renewcommand{\floatpagefraction}{0.8}
%%%%%%%%%%%%%%%%%%%%%%%%%%%%
\newcommand{\ba}{{\bf a}}
\newcommand{\br}{{\bf r}}
\newcommand{\bn}{{\bf n}}
\newcommand{\bq}{{\bf q}}
\newcommand{\bv}{{\bf v}}
\newcommand{\bb}{{\bf b}}
\newcommand{\bee}{{\bf e}}
\newcommand{\bk}{{\bf k}}
%%%%%%%%%%%%%%%%%%%%%%%%
\def\ep{\epsilon}
\def\vep{\varepsilon}
\def\al{\alpha}
\def\be{\beta}
\def\ga{\gamma}
\def\Ga{\Gamma}
\def\la{\lambda}
\def\th{\theta}
\def\de{\delta}
\def\De{\Delta}
\def\si{\sigma}
\def\ti{\tilde}
\def\et{\eta}
\def\pa{\partial}
\def\om{\omega}
\def\ka{\kappa}
\def\Om{\Omega}
\def\fr{\frac}
\def\be{\begin{equation}}
\def\ee{\end{equation}}
\def\bea{\begin{eqnarray}}
\def\eea{\end{eqnarray}}
%%%%%%%%%%%%%%%%%%%%%%%%%%%%%%%%%%%%%%%%%%%%%%%%%%%%%%%%%%%%%%%%%%%%%%%%%%%%%%%%%
\title{Nonlinear Modulation of Multi-Dimensional Lattice Waves}
\author{Guoxiang Huang$^{1,2}$, Vladimir V. Konotop$^{3}$, Hon-Wah Tam$^{4}$
        and Bambi Hu$^{2,5}$ }
\address{ $^{1}$Key Laboratory for Optical and
                Magnetic Resonance Spectroscopy and Department of Physics,
                East China Normal University,
                Shanghai 200062,
                China\\
          $^{2}$Centre for Nonlinear Studies and Department of Physics,
                Hong Kong Baptist University, Hong Kong, China\\
          $^{3}$Departamento de F\'{\i}sica and Centro de F\'{\i}sica
               da Mat\'eria Condensada, Universidade de Lisboa, Complexo
                Interdisciplinar, Avenida Professor
                Gama Pinto, 2, Lisbon P-1649-003, Portugal\\
          $^{4}$Department of Computer Science,
                Hong Kong Baptist University, Hong Kong, China\\
          $^{5}$Department of Physics, University of Houston, Houston
                TX 77204, USA
        }
\maketitle
%%%%%%%%%%%%%%%%%%%%%%%%%%%%%%%%%%%%%%%%%%%%%%%%%%%%%%%%%%%%%%%%%%%%%%%%%%%
\begin{abstract}
The equations governing weakly nonlinear modulations of
$N$-dimensional lattices are considered using a quasi-discrete
multiple-scale approach. It is found that the evolution of a
short wave packet for a lattice system with cubic and quartic
interatomic potentials is governed by generalized
Davey-Stewartson (GDS) equations, which include mean motion
induced by the oscillatory wave packet through cubic interatomic
interaction. The GDS equations derived here are more
general than those known in the theory of water waves
because of the anisotropy inherent in lattices.
Generalized Kadomtsev-Petviashvili equations describing the
evolution of long wavelength acoustic modes in two and three
dimensional lattices are also presented.
Then the  modulational instability of a $N$-dimensional Stokes lattice wave
is discussed based on the $N$-dimensional GDS equations obtained. Finally,
the one- and  two-soliton solutions of two-dimensional GDS equations
are provided by means of  Hirota's bilinear transformation method.
\end{abstract}
%%%%%%%%%%%%%%%%%%%%%%%%%%%%%%%%%%%%%%%%%%%%%%%%%%%%%%%%%%%%%%%%%%%%%%%%%%%%
%
\section{INTRODUCTION}

Since the pioneering work of Fermi, Pasta and Ulam\cite{fer} on
the nonlinear dynamics in lattices, the understanding of the
dynamical localization in ordered, spatially extended discrete
systems has experienced considerable progress. In particular,
one-dimensional(1D) lattice solitons, which are localized
nonlinear excitations due to the balance between nonlinearity and
dispersion, are shown to exist\cite{rem}. Similar
to the cases in fluid physics and nonlinear optics, most of the
analytical approaches on lattice solitons are based on weakly
nonlinear theory. The basic idea of the weakly nonlinear theory is
that linearized lattice equations are assumed to provide a
satisfactory first approximation for those finite-amplitude
disturbances which are, in some sense, sufficiently small.
Successive approximations may then be developed by an asymptotic
expansion in ascending powers of a characteristic wave amplitude.
The weakly nonlinear theory has been shown to be very successful
in revealing many important physical processes, e.\,g. resonant
wave-wave interactions, modulational instability, the formation
of solitons, etc, in a clear-cut way. A very useful method for
the asymptotic expansion is the method of multiple-scales, which
in the case of lattices reduces the system to a set of partial
differential equations for the envelope\,(or amplitude) while the
original system is a set of differential-difference equations,
and usulally cannot be solved exactly. There are two basic advantages
of the multiple scale expansion: (i) it contains a unique explicit
small parameter, and hence is controlable, and (ii) it allows obtaining
solutions in an explicit form. It is well known that, for
a 1D lattice wave with a large spatial extension, the envelope of
the lattice wave is governed by the nonlinear Schr\"{o}dinger
(NLS) equation for a short wavelength packet\cite{tsu} and the
Korteweg-de Vries (KdV) equation for a long wavelength acoustic
mode\cite{kru}.

In recent years, much attention has been paid to coherent
structures in multi-dimensional lattices (see e.g.
Ref.\cite{dru}). In particular we mention a generalization of the
KdV equation in a 2D lattice with only a cubic interatomic
potential, i.\,e. the Kadomtsev-Petviashivili (KP) equation,
derived for a lattice wave traveling in a given
direction\cite{dun} and coupled 2D NLS equations describing
quadratic solitons due to the second harmonic generation in a 2D
lattice of the two-component dipoles \cite{KonMal}. However, to
the best of our knowledge, up to now 2D and 3D generalization of
the NLS equation with a mean motion induced by oscillatory wave
packets in lattice systems (i.e. due to long wavelength acoustic
mode) has not been developed. Meantime such  motion introduces
dramatical changes in the lattice dynamics.

In the present paper, using a quasi-discrete multiple-scale
approach\cite{tsu,hua1,hua2,kon}, we derive generalized
Davey-Stewartson (GDS) equations in multidimensional lattices with
cubic and quartic interatomic potentials. Because of the
anisotropy inherent in lattice systems\,(i.\,e. without
continuous translation and rotation symmetries),
in the case of two dimensions the GDS equations
presented here are more general than those obtained in water
waves\cite{dav}, which are physically isotropic. We also derive a
generalized  KP equation governing the evolution of a long
wavelength acoustic excitation traveling in any direction.

The organization of the paper is as follows. In Sec. II we
formulate the model and deduce the equations for slowly varying
amplitudes in an $N$D lattice. In Sec. III we concentrate on
excitations in a 2D lattice. The dynamical equations in the
long-wavelength limit are presented in Sec. IV.
In Sec. V we discuss modulational instability of a $N$-dimensional
Stokes lattice  wave on the
basis of the GDS equations. Section VI provides
some one- and two-soliton solutions for the 2D  GDS
equations based on Hirota's bilinear transformation  method and
demonstrate the effect of anisotropy on the soliton formation.
The outcomes are summarized in the last section.

\vspace{2mm}

%%%%%%%%%%%%%%%%%%%%%%%%%%%%%%%%%%%
%
\section{Model and asymptotic expansion}

The system we consider is a monoatomic scalar lattice with
nearest-neighbor interatomic interactions. The equations of motion
describing the system are given by
\bea \label{lattice}
 \fr{d^2}{dt^2}u(\bn) =& &
\sum_{j=1}^d J_{2j}[u(\bn+\ba_j)+u(\bn-\ba_j)-2u(\bn)] \nonumber
\\ & &
+\sum_{j=1}^d
J_{3j}\{[u(\bn+\ba_j)-u(\bn)]^2-[u(\bn-\ba_j)-u(\bn)]^2\}
  \nonumber \\ &
&
+\sum_{j=1}^dJ_{4j}\{[u(\bn+\ba_j)-u(\bn)]^3+[u(\bn-\ba_j)-u(\bn)]^3\}.
\eea
Here $u(\bn)$ is the displacement from its equilibrium position of
the particle having the mass $M$ and located at the site
$\bn=\sum_{j=1}^dn_j\ba_j$, $n_j$ being integers, $\ba_j$ being
the lattice vectors, and $d$ being the dimension of the lattice,
$J_{\alpha j}=K_{\alpha,j}/M\,(\alpha=2,\,3,\,4)$,  $K_{2,j},\,
K_{3,j}$ and $K_{4,j}$ are harmonic, cubic and quartic
nearest-neighbor force constants, respectively. Notice that the
anisotropy of the lattice is included in the consideration (i.e.
in a generic case $K_{\alpha,j}\neq K_{\alpha,i}$ for $i\neq j$).
We include the cubic potential here since most of
realistic interatomic potentials\,(such as the potentials of
Born-Mayer-Coulomb, Lennard-Jones, Morse, Toda, etc.) display
strong cubic nonlinearity\,(i.\,e. $J_{\alpha,3}\neq
0$)\cite{hua1,hua2}. In the most direct physical applications
(namely, to atomic crystals) the dimension $d$ can be either 2 or
3, although more formal lattices with $d$ being bigger than 3 are
available. In the present section we deal with the last, more
general, case.

In order to investigate weakly nonlinear modulation of a lattice
wave packet we use the quasi-discrete multiple-scale
method\cite{tsu,hua1,hua2,kon}
to derive the envelope equations describing the
development of the modulation of the packet along the line of
Davey and Stewartson for water waves\cite{dav}. Namely, we set
\be
\label{expan1} u(\bn)=\sum_{\nu=1}\ep^\nu u_\nu(\br,\tau;\phi(\bn,
t)) \ee
with
\bea \label{expan2} \br=\ep (\bn-\bv t),\qquad  \phi(\bn,
t)={\bf q \cdot n}-\om t, \eea
where $\ep$ is a formal small parameter representing the relative
amplitude of the excitation,  ${\bf q}$ is the wave vector: ${\bf
q}=\sum_{j=1}^dq_j\bb_j$, $\bb_j$ being the vectors of the
reciprocal lattice: $\bb_i\cdot\ba_j=\delta_{ji}$  and $\om$  is
the frequency of the respective harmonic.  The constant vector
$\bv$ as well as the link between $\omega$ and ${\bf q}$, i.e.
the dispersion relation, are to be determined by solvability
conditions.

There are some comments to be made here. In a generic case the
entries of the expansion (\ref{expan1}) depend on the whole
hierarchy of "slow" variables, i.e. one should consider the set of
variables  $\{\br_\nu,t_\nu\}$ ($\nu=1,2,...$) where
$\br_\nu=\ep^\nu (\bn-\bv_\nu t)$ and $t_\nu=\ep^\nu t$, which
are regarded as independent. In the case one is interested in the
effect of quadratic and cubic nonlinearity only the scales up to
$\br_2$ and $t_2$ turn out to be relevant. In the present paper we
restrict our consideration  to the solutions independent on
$\br_2$ and this a reason of introducing only the "lowest order"
slow variables $\br=\br_1$ and $\tau=t_2$.

Substituting (\ref{expan1}) and (\ref{expan2}) into
Eq.(\ref{lattice}) and equating the coefficients of the same
powers of $\ep$, we obtain the hierarchy of equations as follows
\be \label{system} Lu_\nu\equiv \omega^2 \frac{\partial^2
u_\nu}{\partial\phi^2}-\sum_jJ_{2j}\left(u_\nu^{(j)}+u_\nu^{(-j)}\right)=M_\nu\,,
\qquad \nu=1,2,\ldots\ee
Here  $u_\nu^{(\pm j)}\equiv u_\nu(\br,\tau;\phi(\bn, t)\pm
q_j)-u_\nu(\br,\tau;\phi(\bn, t)), $
\begin{mathletters}
\begin{equation}
 \label{m1}
  M_1=0,
\end{equation}
\begin{equation}
\label{m2}
  M_2=-2\omega(\bv\cdot\nabla)\frac{\partial
 u_1}{\partial\phi}+\sum_jJ_{2j}a_j\frac{\partial}{\partial x_j}
  (u_1^{(j)}-u_1^{(-j)})+
 \sum_jJ_{3j}[\left(u_1^{(j)}\right)^2-\left(u_1^{(-j)}\right)^2],
 \end{equation}
 \begin{eqnarray}
 \label{m3}
M_3=-(\bv\cdot\nabla)^2u_1 -2\omega(\bv\cdot\nabla)\frac{\partial
u_2}{\partial \phi}+2\omega\frac{\partial^2 u_1}{\partial
\phi\partial\tau} +\sum_jJ_{2j}a_j\frac{\partial}{\partial
x_j}(u_2^{(j)}-u_2^{(-j)})
\nonumber \\
+  \sum_j\frac{J_{2j}}{2}\left(a_j\frac{\partial}{\partial
x_j}\right)^2 \left(u_1^{(j)}+u_1^{(-j)}+2u_1\right)+
 2\sum_jJ_{3j}\left[(u_2^{(j)}u_1^{(j)}-u_2^{(-j)}u_1^{(-j)})
  \right. \nonumber \\ \left.
+ u_1^{(j)}a_j\frac{\partial}{\partial
x_j}\left(u_1^{(j)}+u_1\right)-
 u_1^{(-j)}a_j\frac{\partial}{\partial x_j}\left(u_1^{(-j)}+u_1\right)\right]+
 \sum_jJ_{4j}\left[\left(u_1^{(j)}\right)^3+\left(u_1^{(-j)}\right)^3
 \right],
\end{eqnarray}
\end{mathletters}
$\nabla\equiv\partial/\partial\br$, $a_j=|{\bf a}_j|$, and $x_m$
is the $m$th coordinate of the vector ${\bf r}$,
$\br=\sum_mx_m {\bf a_m}/a_m$.

For further consideration we have to specify the effect we are
looking for and this will determine the form of
lowest-order\,($j=1$) solution of Eq.(\ref{system}). Namely we will
be interested in the weakly nonlinear modulation of a lattice
wave originated by the interaction between a
long wave-length acoustic mode and a high frequency mode. Thus we choose
\bea \label{firstorder}
 u_1=A_{0}(\br,\tau)
  +\{ A_{1}(\br,\tau)\exp[i\phi(\bn, t)]+c.c.\},
\eea
where the real function $A_0$ stands for a mean motion induced by
the oscillatory wave packet, which has the complex envelope
function $A_1$, and $c.c.$ denotes corresponding complex
conjugate term. Then
\[
u_1^{(\pm j)}=[\exp(\pm i q_j) -1]A_1e^{i\phi({\bf n}, t)}+ c.c.,
\]
and in the first order [see Eqs. (\ref{system}), (\ref{m1})] we
immediately arrive at the dispersion relation of the underline
linear lattice
\begin{equation}
\label{dispersion} \om^2\equiv[\om(\bq)]^2=2\sum_jJ_{2j}(1-\cos
q_j).
\end{equation}

Next we take into account that
\begin{equation}
\label{grvel}
 \bv_g\equiv\frac{d\omega}{d \bq}=
\frac{1}{\om}\sum_jJ_{2j}\sin(q_j)\ba_j/a_j,
\end{equation}
which is the group velocity of the linear wave. Then, subject to
assumption (\ref{firstorder}) the second order equation of system
(\ref{system}) takes the form
\begin{eqnarray}
\label{sys2}
Lu_2=2i\om\left\{\left((\bv_g-\bv)\cdot\nabla\right)\left(
A_1e^{i\phi}-\bar{A}_1e^{-i\phi}\right)+\chi^{(2)}\left(
A_1^2e^{2i\phi}-\bar{A}_1^2e^{-2i\phi}\right)\right\}
\end{eqnarray}
where
\begin{equation}
\label{chi2} \chi^{(2)}= \sum_m \frac{J_{3m}}{\omega}(\cos
q_m-1)\sin q_m
\end{equation}
 is the effective quadratic nonlinearity.

The solvability condition for the system (\ref{sys2}) (in other
words the conditions of the absence of secular terms in $u_2$)
means the orthogonality of the right hand side of Eq.(\ref{sys2}) to
the kernel of the operator $L$, i.e. to  (\ref{firstorder}). Hence
the r.h.s. of Eq.(\ref{sys2}) must do not contain the terms
proportional to $\exp(\pm i\phi)$ and we conclude that
$\bv=\bv_g$, i.e. $\bv$ introduced in (\ref{expan2}) is nothing
but the group velocity of the high frequency carrier wave. Next we
can look for the solution $u_2$ (it must be orthogonal to the
first order approximation, i.e. to the kernel of the operator $L$)
in a form of the expansion over the eigenfunctions of the operator
$L$. Having done this one ensures that the only nonzero term of
such an expansion is given by
\be \label{secondorder}
 u_2=i\alpha A_1^2 \exp (2i\phi)+c.c., \qquad
\alpha=-\frac{2\omega\chi^{(2)}}{4[\omega(\bq)]^2-[\omega(2\bq)]^2}.
\ee

Formula (\ref{secondorder}) is valid unless the condition $
\omega(2\bq)=2\omega(\bq)$
is satisfied. As it is evident this is the condition of the
resonant second harmonic generation \cite{kon,gon}. It can be satisfied
in a
lattice with a complex cell, but it is not difficult to ensure that
$\omega(2\bq)\neq 2\omega(\bq)$ for all ${\bf q}$ in a monoatomic
lattice with the
nearest neighbor interactions.

Passing to the third order of the multiple scale expansion we
introduce the (symmetric) group velocity dispersion tensor (GVDT)
by the formula ($v_j=\partial \omega/\partial q_j$)
\begin{equation}
\label{GVDT} \omega_{ij}\equiv
 \frac{1}{\omega}[J_{2j}\cos (q_j)a_ia_j\delta_{ij}-v_iv_j],
\end{equation}
and the (symmetric) effective GVDT $\Omega_{ij}$
\begin{equation}
\label{EGVDT} \Omega_{ij}\equiv
\frac{1}{\omega}[J_{2j}a_ia_j\delta_{ij}-v_iv_j].
\end{equation}

Then the solvability condition for the third order terms gives
rise to the closed system of equations for $A_0$ and $A_1$:
\begin{equation}
\label{eqa0}
 \sum_{l,m}\Omega_{lm}\frac{\partial^2}{\partial
x_l\partial x_m}A_0=-2\sum_m\delta_m\frac{\partial}{\partial
x_m}|A_1|^2,
\end{equation}
\begin{equation}
\label{eqa1}
 i\frac{\partial A_1}{\partial\tau}+\frac{1}{2}\sum_{l,m}\omega_{lm}
 \frac{\partial^2}{\partial
x_l\partial x_m}A_1=\chi |A_1|^2A_1+A_1\sum_m\delta_m
\frac{\partial }{\partial x_m}A_0,
\end{equation}
where
\bea
& & \delta_m=\frac{2a_m}{\omega}J_{3m}(1-\cos q_m),\\
& & \chi=\frac{2}{\omega}\sum_m[2\alpha J_{3m}(1-\cos q_m)\sin
q_m+ 3 J_{4m}(1-\cos q_m)^2].
\eea
We call Eqs.(14) and (15) the $N$D GDS equations.

%%%%%%%%%%%%%%%%%%%%%%%%%%%%%%%%%%%%%%%%%%%%%%%%%%%%%%%%%%
\section{Generalized Davey-Stewartsons}

Let us now focus our attention on a special case of a 2D lattice
(i.e. ${\bf r}=(x_1,x_2)$). For the sake of simplicity the
lattice will be considered symmetric, $J_{\alpha,j}=J_\alpha$
($\alpha=2,3,4$ and $j=1,2$) and orthogonal: $\ba_1\cdot\ba_2=0$,
with the lattice constant equal to unity, $|a_j|=1$.
In order to
diagonalize the effective GVDT $\Omega_{lm}$ in a general case we
rotate the original Cartesian system (with the coordinate basis
(1, 0) and (0, 1)\,) to a new one with the coordinate basis ${\bf
e}_1=(\la_1, \la_2)$ and ${\bf e}_2=(-\la_2, \la_1)$, where
\be \label{lambda} \lambda_j= \frac{v_j}{v_g} = \fr{\sin
q_j}{\sqrt{\sin^2 q_1+\sin^2 q_2}}. \ee
and $v_j$ is the $j$-th component of the group velocity defined
in (\ref{grvel}) (as it is evident ${\bf e}_i\cdot{\bf e}_j=\delta_{ij}$).
In this way one
of the directions of the new basis, namely ${\bf e}_1$ coincides
with the direction of the group velocity of the carrier wave, i.
e. $\bv_g=v_g{\bf e}_1$ the other direction is orthogonal to it.  As a result, $x_1$ and $x_2$ in Eqs.(14)
and (15) take the form $x_1=\ep (n_1-\la_1 v_g t)$ and   $x_2=\ep
(n_2-\la_2 v_g t)$, and envelope equations  (\ref{eqa0}) and
(\ref{eqa1}) are reduced to
\begin{figure}[h]
\vspace{0.5 true cm}
\includegraphics{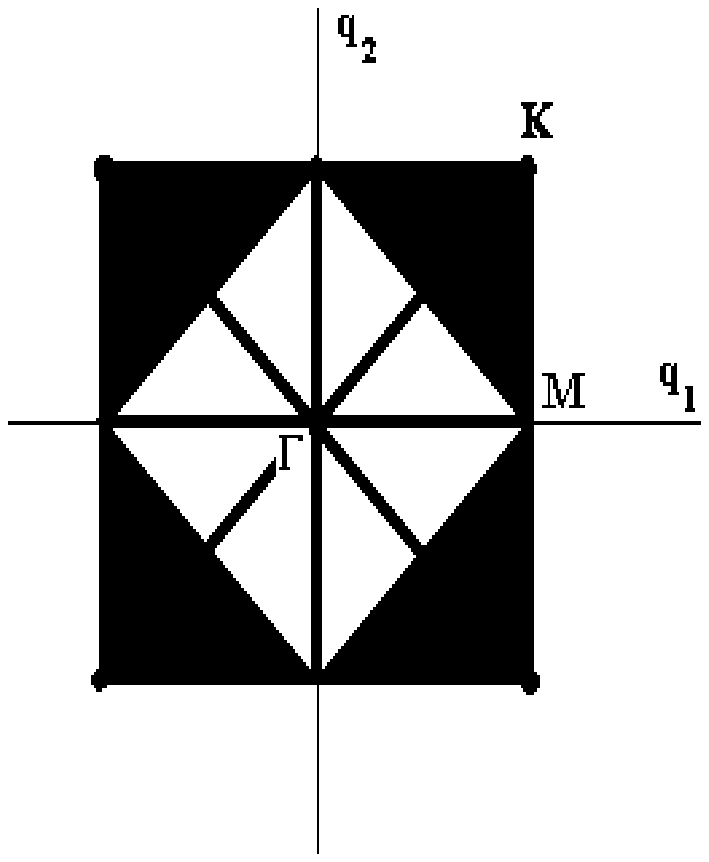}
\vspace{-1 true cm}
\caption{The first Brillouin zone for the 2D quadratic lattice.
The filled in and empty polygons correspond to the operator ${\cal L}$ (it is
defined by (\ref{Loper})\,) of the elliptic and hyperbolic types, respectively.
Along the intervals shown by the bold lines (i.e. in the directions [$100$],
[$010$], [$110$], and [$1\bar{1}0$]) indicates the system (\ref{DS11}),
(\ref{DS12}) is reduced to the conventional DSII equation.}
\label{figone}
\end{figure}
\begin{eqnarray} \label{DS11} \al_{11}\fr{\pa^2 A_0}{\pa
\xi^2}+\al_{22}\fr{\pa^2 A_0}{\pa \eta^2}
    =-2\left( \beta_1 \fr{\pa}{\pa\xi}+ \beta_2 \fr{\pa}{\pa\eta}\right)
     |A_1|^2,
\end{eqnarray}
 \begin{eqnarray}
     \label{DS12}
i\fr{\pa A_1}{\pa \tau}+ {\cal L} A_1
=A_1\left(\beta_1 \fr{\pa}{\pa \xi}+\beta_2 \fr{\pa}{\pa \eta}\right)
     A_0+\chi |A_1|^2 A_1,
\end{eqnarray}
where
\begin{equation}
\label{Loper}
{\cal L}=\ga_{11}\fr{\pa^2 }{\pa \xi^2}
    +\ga_{22}\fr{\pa^2 }{\pa \eta^2}
    +\ga_{12}\fr{\pa^2 }{\pa\xi \pa\eta},
\end{equation}
\bea
& & \xi=\br\cdot{\bf e}_1=\lambda_1x_1+\lambda_2x_2=\ep (\la_1 n_1
        +\la_2 n_2-v_g t),\\
& & \eta=\br\cdot{\bf e}_2=-\lambda_2 x_1+\lambda_1 x_2=\ep (-\la_2 n_1
        +\la_1 n_2),\\
& & \al_{11}=\frac{1}{\omega}(J_2 -v_g^2),\,\,\,\,\al_{22}=\frac{J_2}{\omega},\\
& & \beta_1=\frac{2J_3}{\omega}[\la_1 (1-\cos q_1)
    +\la_2 (1-\cos q_2)],\,\,\,\,\beta_2=\frac{2J_3}{\omega}[\la_1 (1-\cos q_2)
    -\la_2 (1-\cos q_1)],\\
& & \ga_{11}=\fr{1}{2\om}[-v_g^2+J_2(\la_1^2\cos q_1
    +\la_2^2\cos q_2)],\,\,\,\,\ga_{22}=\fr{J_2}{2\om}(\la_2^2\cos q_1
    +\la_1^2\cos q_2),\\
& & \ga_{12}=\fr{J_2}{\om}\la_1 \la_2 (\cos q_2-\cos q_1),\\
& & \chi=\fr{2}{\om}\left\{ 2J_3\alpha[\sin q_1 (1-\cos q_1)
           +\sin q_2 (1-\cos q_2)]+3J_4[(1-\cos q_1)^2+(1-\cos q_2)^2 ]
           \right\},\\
& & \alpha =\fr{4J_3[\sin q_1(1-\cos q_1)+\sin q_2(1-\cos q_2)]}
      {4[\om({\bf q})]^2-[\om(2{\bf q})]^2}.
\eea

Equations (\ref{DS11}) and (\ref{DS12}) represent a generalized
form of the conventional DS equations. They include the dispersion,
diffraction and nonlinearity of the system. One of their
important features is that there exists a coupling between the mean
field\,(denoted by $A_0$) and the envelope of the carrier
wave\,(denoted by $A_1$). The mean field $A_0$ generates a strain
field in the system. If $J_3=0$, a case for a symmetric
interatomic potential, we have $A_0=0$ thus the mean motion and
hence the strain field vanishes. Another important feature for
Eqs.(\ref{DS11}) and (\ref{DS12}) is their property of anisotropy.
For different
wave vector ${\bf q}=(q_1, q_2)$, the coefficients of the
equations take different values and some of these coefficients
may become vanishing for some particular directions of ${\bf q}$.

The conventional DS equations were derived firstly in surface
water waves\cite{dav} and now are a well-known 2D soliton model
in the soliton
theory\cite{abl1}.  Note that for water waves, the system is
isotropic (i.\,e. it possesses a continuous rotation symmetry).
The envelope equations are the same for all propagating
directions of the waves and hence the coefficients appearing in
the equations are independent on $q_1$ and $q_2$ and
correspondingly $\beta_2$, and $\ga_{12}$ vanish\cite{dav}\,(see
also Ref.\cite{abl2}).   However, {\it for the lattice system the
modulating equations take a more general form because the lattice
is anisotropic\,(without the continuous rotation symmetry)}. We
 mention that although the coefficients $\alpha_{ij}$ are
both positive, signs of the coefficients $\gamma_{ij}$ may change
depending on the choice of the wave-vector in the first Brillouin
zone.

We now discuss several particular cases for the 2D GDS equations derived
above. In the following circumstances\,(i.\,e. in some special points
and lines of the Brillouin zone, see Fig. 1) the 2D GDS equations reduce to
the conventional  DS equations:
\begin{enumerate}
\item $q_1q_2= 0$ (then $\lambda_1\lambda_2=0$, and
$\beta_2=\ga_{12}=0$)
\item $q_1=q_2=q$ (then $\lambda_1=\lambda_2=2^{-1/2}$ and
$\beta_2=\ga_{12}=0$)
\item $q_1=-q_2=q$ (then $\lambda_1=-\lambda_2=2^{-1/2}$ and
$\beta_1=\ga_{12}=0$)
\end{enumerate}
More precisely, since $\alpha_{11}>0$ and $\alpha_{22}>0$ for any
${\bf q}$,  at $\gamma_{11}\gamma_{22}<0$ Eqs. (\ref{DS11}) and (\ref{DS12})
can be classified as DSII equations, while for
$\gamma_{11}\gamma_{22}>0$ they form a dynamical system that can be
identified neither with DSI nor DSII equations appearing in the
theory of water waves (see e.g. \cite{abl1}).

In the case of pure quadratic potential, $J_3=0$, we have that
the evolution equations for $A_1$ and $A_0$ are decoupled. Then
the GDS equations reduce to a generalized 2D NLS
equation\,(i.\,e. the NLS equation plus a cross-derivative term
$\pa^2 A_1/(\pa \xi\pa\eta)$). Finally, if $q_2=0$ and
$\pa/\pa\eta=0$, the 2D GDS equations (\ref{DS11}) and (\ref{DS12})
recover the
envelope equations derived in Ref.\cite{hua1,kon}, which gives
rise to standard 1D lattice solitons\cite{hua1,kon}.

In the 3D case, Eqs. (\ref{DS11}) and (\ref{DS12}) are replaced by
the {\it 3D GDS equations}:
\bea
& &  \label{3dds1}
    \al_{11}^{\prime}\fr{\pa^2 A_0}{\pa \xi^2}+\al_{22}^{\prime}\fr{\pa^2
    A_0}{\pa \eta^2}
    +\al_{33}^{\prime}\fr{\pa^2 A_0}{\pa \zeta^2}
    =-2\left( \beta_1^{\prime} \fr{\pa}{\pa\xi}+ \beta_2^{\prime}
    \fr{\pa}{\pa\eta}+\beta_3^{\prime} \fr{\pa}{\pa\zeta}\right)
     |A_1|^2,\\
& & \label{3dds2}
    i\fr{\pa A_1}{\pa \tau}+\ga_{11}^{\prime}\fr{\pa^2 A_1}{\pa \xi^2}
    +\ga_{22}^{\prime}\fr{\pa^2 A_1}{\pa \eta^2}+\ga_{33}^{\prime}
    \fr{\pa^2 A_1}{\pa \zeta^2}
    +\left(\ga_{12}^{\prime}\fr{\pa^2 }{\pa\xi \pa\eta}+\ga_{23}^{\prime}
    \fr{\pa^2 }{\pa\eta\pa\zeta}
    +\ga_{31}^{\prime}\fr{\pa^2 }{\pa\zeta\pa\xi}\right)A_1\nonumber\\
& & =A_1\left(\beta_1^{\prime} \fr{\pa}{\pa \xi}
    +\beta_2^{\prime} \fr{\pa}{\pa \eta}+\beta_3^{\prime} \fr{\pa}{\pa \zeta}
    \right)A_0+\chi^{\prime} |A_1|^2 A_1,
\eea
where $\al_{jj}^{\prime}$, $\beta_j^{\prime}$,
$\ga_{ij}^{\prime}$\,($j=1,\,2,\,3)$
and $\chi^{\prime}$ are constants dependent on ${\bf q}=(q_1,q_2,q_3)$
and the parameters of the system, which are not needed here and not
written down expliciltly. The definitions of $\xi, \eta$ and $\zeta$
are given by
\bea
& & \label{xi1} \xi=\ep (\la_1 n_1+\la_2 n_2+\la_3 n_3-v_g t), \\
& & \label{xi2} \eta=\ep (-\la_2 n_1+\la_1 n_2),\\
& & \label{xi3} \zeta=\ep (-\la_3 n_1+\la_1 n_3),
 \eea
 where
$\om$ and $\la_j$ are defined by (\ref{dispersion}) and
(\ref{lambda}).

%%%%%%%%%%%%%%%%%%%%%%%%%%%%%%%%%%%%%%%%%%%%%%%%%%%%%%%
\section{Long-wavelength limit}

Note that the envelope equations (\ref{eqa0}) and (\ref{eqa1})
are invalid for ${\bf q}=0$
since in this case there is a divergence in their coefficients.
From the physical point of view this happens because vanishing
${\bf q}$ corresponds to a long wavelength acoustic mode in the lattice.
In this case a different asymptotic expansion must be used to obtain
divergence-free envelope equations. For simplicity we consider the case of a symmetric 2D square lattice.
In this situation the asymptotic expansion (\ref{expan1})  must be replaced by
\be\label{expan4}
u=u_{0}+\ep u_{1}+\ep^2 u_2+\cdots,
\ee
with
\bea
& & u_{\nu}=u_{\nu}(\xi,\eta,\tau)\,\,\,\nu=0,\,1,\,2,\cdots, \\
& & \xi=\ep (\la_1 n_1+\la_2 n_2-ct),\\
& & \eta=\ep^2 (-\la_2 n_1+\la_1 n_2), \\
& & \tau=\ep^3 t,
\eea
where  $c=\sqrt{J_2}$ is the speed of sound,  $\la_l\,(l=1,\,2)$ are determined by the
solvability conditions required at $O(\ep^2)$-order.
A solvability condition in the forth-order of the expansion yields the
{\it generalized KP equation}:
\be
\label{eq:gkp}
\fr{\pa}{\pa\xi}\left [ \fr{\pa v}{\pa
\tau}+\fr{c}{24}(\la_1^4+\la_2^4)\fr{\pa^3 v}{\pa \xi^3}
+\fr{\tilde{J_3}}{c}(\la_1^3+\la_2^3) v \fr{\pa v}{\pa \xi}
+\fr{3J_4}{2c}(\la_1^4+\la_2^4) v^2 \fr{\pa v}{\pa \xi} \right]
+\fr{c}{2}\fr{\pa^2 v}{\pa \eta^2}=0,
\ee
where $v=\pa u_0/\pa \xi$. In deriving
Eq.(\ref{eq:gkp}) we have assumed that $J_3=\ep \tilde{J_3}$ with $\tilde{J_3}$
of order unity.   The parameters $\la_l(l=1, 2)$ are direction-dependent
and we find that their values can be obtained by using (\ref{lambda})
but taking the limit ${\bf q}\rightarrow 0$.
Thus the values of the coefficients in
Eq.(\ref{eq:gkp}) are dependent on the ways of ${\bf q}$ approaching zero.
For instances
\begin{enumerate}
\item $\la_1=1$, $\la_2=0$,  if $q_2=0$, $q_1\rightarrow 0$;
\item $\la_1=\la_2=1/\sqrt{2}$, if $q_1=q_2=q\rightarrow 0$.
\end{enumerate}
The reason for appearing different values
of the coefficients corresponding to different directions is also due to the
anisotropy of the system. It is easy to see that the KP equation obtained in
Ref.\cite{dun} is our particular case with
quartic nonlinearity being absent
(i.\,e. $J_4=0$). Eq.(\ref{eq:gkp}) admits solitary wave
solutions\cite{abl1}.

It is relevant to mention here that the coefficient of the term
$\partial^2v/\partial\eta^2$ is positive which means that the
line (i.e. $\eta$-independent) solitons of Eq. (\ref{eq:gkp}) is
stable while this equation does not admit any kind of lump (i.e.
decaying when $\xi^2+\eta^2\to 0$) solution.

In the same way, in 3D case Eq.(\ref{eq:gkp}) is generalized to
\be\label{3gkp}
\fr{\pa}{\pa\xi}\left[
    \fr{\pa v}{\pa \tau}+a_1\fr{\pa^3 v}{\pa \xi^3}
    +a_2 v \fr{\pa v}{\pa \xi}
    +a_3 v^2 \fr{\pa v}{\pa \xi}\right] \nonumber\\
    +a_4 \fr{\pa^2 v}{\pa \eta^2}
    +a_5 \fr{\pa^2 v}{\pa \zeta^2}=0,
\ee
 where $\xi, \eta$ and $\zeta$ are the same as (\ref{xi1})-(\ref{xi3}).
$a_l\,(l=1,\,2,\,3,\,4,\,5)$ are real constants dependent on
$\la_j\,(j=1,\,2,\,3)$(given by \ref{lambda}) with ${\bf q}\rightarrow 0$.
%
%
%
%%%%%%%%%%%%%%%%%%%%%%%%%%%%%%%%
\section{Modulational instability of a plane lattice wave
            with a mean motion}
%%%%%%%%%%%%%%%%%%%%%%%%%%%%%%%%%

In recent years, the use of nonlinear envelope(or amplitude) equations for
studying the stability of patterns and waves in systems in and
outside of equilibrium is widely employed\cite{stu,cro}. The modulational
stability of a plane water wave\,(e. g. a uniform Stokes wave) was analyzed by
Davey and Stewartson based on the DS equations they derived\cite{dav}.
In the same way the $N$D GDS
equations (\ref{eqa0}) and (\ref{eqa1}) obtained here can be used to study the
modulational stability of a uniform Stokes lattice wave in $N$ dimensions.
A Stokes lattice wave  here means a linear plane lattice wave with the
wave vector ${\bf q}$.

Note that the uniform vibrating solution of Eqs.(\ref{eqa0})
and (\ref{eqa1}) reads
\be \label{unif}
A_0=0, \,\,\,\,\,\, A_1=U_0\exp (-i\Omega \tau),
\ee
which, when incoperating the carrier wave (see (\ref{firstorder})\,), corresponds
a plane lattice wave with the wave vector ${\bf q}$ and the frequency
$\om({\bf q})+\Omega$ excited in the system, where $U_0$ is a constant
and $\Omega=\chi U_0^2$. Assume that a perturbation is added into the
uniform vibreting solution (\ref{unif}), i.\,e.
\bea
& & \label{pert1}A_0(x_1,\,x_2,\,...,\,\tau)
    =\hat{\ka}_{+}\exp (i\sum_m Q_m x_m)
    +\hat{\ka}_{-}\exp (-i\sum_m Q_m x_m),\\
& & \label{pert2}A_1(x_1,\,x_2,\,...,\,\tau)
    =U_0\exp (-i\Omega \tau)\,\left [1+
    \hat{\vep}_{+}\exp (i\sum_m Q_m x_m)
    +\hat{\vep}_{-}\exp (-i\sum_m Q_m x_m)\,\right],
\eea
with
$\hat{\ka}_{\pm}(\tau)=\ka_{\pm}(0) \exp [(\si_R\pm i\si_I)\tau]$ and
$\hat{\vep}_{\pm}(\tau)=\vep_{\pm}(0) \exp [(\si_R\pm i\si_I)\tau]$,
where ${\bf Q}$=$(Q_1,\,Q_2,\,...,\,Q_N)$ and $\si_I$=$\si_I({\bf Q})$
are respectively the wave vector and frequency of the perturbation,
$\si_R=\si_R({\bf Q})$ denotes the growth rate of the perturbation,
$\ka_{\pm}(0)$ and $\vep_{\pm}(0)$ are small constants with the condition
$\hat{\ka}_{-}^{*}(0)=\ka_{+}(0)$ because of the reality of $A_0$.
Substituting (\ref{pert1}) and (\ref{pert2}) into the Eqs.(\ref{eqa0}) and
(\ref{eqa1}) we obtain
a set of linear equations on $\ka_{\pm}(0)$ and $\vep_{\pm}(0)$:
\bea
& & \label{modueq1} -(\al_{11}Q_1^2+\al_{22}Q_2^2)\ka_{+}(0)+2iU_0^2(\beta_1
    Q_1+\beta_2Q_2)[\vep_{+}(0)+\vep_{-}^{*}(0)]=0,\\
& & \label{modueq2} (\al_{11}Q_1^2+\al_{22}Q_2^2)\ka_{-}(0)+2iU_0^2(\beta_1
    Q_1+\beta_2Q_2)[\vep_{-}(0)+\vep_{+}^{*}(0)]=0,\\
& & \label{modueq3} (\Omega+i\si-\ga_{11}Q_1^2-\ga_{22}Q_2^2-\ga_{12}Q_1Q_2
    -2\chi U_0^2]\vep_+(0)-\chi U_0^2\vep_{-}^*(0)-i(\beta_1Q_1
    +\beta_2Q_2)\ka_{+}(0)=0,\\
& & \label{modueq4} (\Omega+i\si^*-\ga_{11}Q_1^2-\ga_{22}Q_2^2-\ga_{12}Q_1Q_2
    -2\chi U_0^2]\vep_-(0)-\chi U_0^2\vep_{+}^*(0)
    +i(\beta_1Q_1+\beta_2Q_2)\ka_{-}(0)=0,
\eea
where $\sigma=\sigma_R+i\sigma_I$.
A solvability condition of Eqs.(\ref{modueq1}-\ref{modueq4}) results in
\be \label{growth}
(\si_R+i\si_I)^2=\left(\sum_{l,m} \om_{lm}Q_l Q_m\right )
\left\{ U_0^2 \left [ -\chi+\fr{2\left(\sum_m \delta_m Q_m\right)^2}
{\sum_{l,m} \Omega_{lm} Q_lQ_m}\right]
-\fr{1}{4}\sum_{l,m} \om_{lm}Q_l Q_m\right\}.
\ee
Note that the right side of Eq.(\ref{growth}) is real. Thus when
\be \label{condi1}
\left(\sum_{l,m} \om_{lm}Q_l Q_m\right )
\left\{ U_0^2 \left [ -\chi+\fr{2\left(\sum_m \delta_m Q_m\right)^2}
{\sum_{l,m} \Omega_{lm} Q_lQ_m}\right]
-\fr{1}{4}\sum_{l,m} \om_{lm}Q_l Q_m\right\}>0,
\ee
one has $\si_I=0$. As a result if the condition (\ref{condi1}) is
satisfied we have the growth rate
\be \label{grow1}
\si_R=\pm \left\{ \left(\sum_{l,m} \om_{lm}Q_l Q_m\right )
\left\{ U_0^2 \left [ -\chi+\fr{2\left(\sum_m \delta_m Q_m\right)^2}
{\sum_{l,m} \Omega_{lm} Q_lQ_m}\right]
-\fr{1}{4}\sum_{l,m} \om_{lm}Q_l Q_m\right\}\,\right\}^{1/2}.
\ee
Thus one always has a positive $\si_R$ branch if the condition
({\ref{condi1}) is satisfied. In this case the perturbation grows
exponentially and hence the unifrom vibrating solution (\ref{unif}) is
modulationally  unstable.

For the 2D GDS equations (\ref{DS11}) and (\ref{DS12}), the condition of the modulational
instability (\ref{condi1}) reads
\bea
& & (\ga_{11}Q_1^2+\ga_{22}Q_2^2+\ga_{12}Q_1 Q_2)\cdot\nonumber\\
& &  \label{condi2} \left\{ U_0^2
\left[ -\chi+\fr{2(\beta_1 Q_1
    +\beta_2 Q_2)^2}{\al_{11}Q_1^2+\al_{22}Q_2^2} \right]
    -\fr{1}{2}(\ga_{11}Q_1^2+\ga_{22}Q_2^2+\ga_{12}Q_1 Q_2)\right\}>0.
\eea
Thus due to the anisotropy of the lattice\,(i.\,e. $\beta_2\gamma_{12}\neq
0)$ the criterion (\ref{condi2}) gives much richer behavior for the
stability of the Stokes wave than that in isotropic systems\,(e.\,g.
water waves). In particular, for a given Stokes lattice wave there exist two
(or may be four depending on the Stokes lattice wave) wave vectors ${\bf Q}$
for which the instability evolves with the biggest incriment.
This phenomenon reminds the so-called strengthing of inhomogeneities,
known in the theory of beam propagaion in Kerr medium \cite{kerr}.
There is however an essential difference originated by the anisotropy:
the biggest exponent is characterized by the amplitude of the value of
the wave vector and also by the lattice direction. The position of the
points providing the largest incriment depends on the choice of the wave
vector of the Stokes lattice wave.

The outcome of this type of instabilty may
result in the fromation of solitons\cite{rem} or the appearence of
homoclinic
structures(see Sec.\,3.3 of Ref.\cite{abl1}).

%
%%%%%%%%%%%%%%%%%%%%%%%%%%%%%%%%%%%%%%%%%%%
%
\section{Soliton solutions}
%
%%%%%%%%%%%%%%%%%%%%%%%%%%%

We now consider the soliton solutions of the nonlinear evolution
equations derived above. Taking 2D GDS equations (\ref{DS11}) and (\ref{DS12}) as an example,
to obtain the soliton solutions we employ Hirota's bilinear transformation
method,
an ingenious technique of finding exact multi-soliton solitons for nonlinear
evolution equations\cite{hir,mat}.
%
% Taking $(R, U)$=$\ep (A_0, A_1)$ and noting
%that $(\xi, \eta, \tau)$=$(\ep x, \ep y, \ep^2 t)$ with $x=\la_1 n_1+\la_2 n_2
%-v_g t$ and $y=-\la_2 n_1+\la_1 n_2$, the 2D
%GDS equations (\ref{DS11})  and  (\ref{DS12}) can be written into the form
%\bea
%& & \label{DS111}\al_{11}\fr{\pa^2 R}{\pa
%    x^2}+\al_{22}\fr{\pa^2 R}{\pa y^2}
%    =-2\left( \beta_1 \fr{\pa}{\pa x}+ \beta_2 \fr{\pa}{\pa y}\right)
%     |U|^2,\\
%& & \label{DS112}i\fr{\pa U}{\pa t}+\ga_{11}\fr{\pa^2 U}{\pa x^2}
%    +\ga_{22}\fr{\pa^2 U}{\pa y^2}
%    +\ga_{12}\fr{\pa^2 U}{\pa x \pa y}
%    =U\left(\beta_1 \fr{\pa}{\pa x}+\beta_2 \fr{\pa}{\pa y}\right)
%     R+\chi |U|^2 U.
%\eea
Introducing the dependent variable transformation
\be
\label{bili}
A_0=-4\left( \beta_1\fr{\pa}{\pa \xi}+\beta_2\fr{\pa}{\pa \eta}\right)\log F,
   \,\,\,\,\,\, A_1=G/F
\ee
with $F$\,(real) and $G$\,(complex) being the functions of $\tau$, $\xi$ and $\eta$,
Eqs.(\ref{DS11}) and (\ref{DS12}) are transformed into the following bilinear form
\bea
& & \label{form1}(\al_{11}D_{\xi}^{2}+\al_{22}D_{\eta}^{2})F\cdot F=|G|^2,\\
& & \label{form2}(iD_{\tau}+\ga_{11}D_{\xi}^2+\ga_{22}D_{\eta}^2+\ga_{12}D_{\xi}D_{\eta})G\cdot F=0,\\
& & \label{form3}[(\ga_{11}-2\beta_1^2)D_{\xi}^2+(\ga_{22}-2\beta_2^2)D_{\eta}^2+
    +(\ga_{12}-4\beta_1 \beta_2)D_{\xi}D_{\eta}]F\cdot F+\chi |G|^2=0,
\eea
where $D_\tau$, $D_\xi$ and $D_\eta$ are Hirota's bilinear operators defined
by\cite{hir,mat}
\bea
& & D_{\xi}^m D_{\eta}^n D_{\tau}^p G\cdot F \nonumber\\
& & \label{defi}
    \equiv\left(\fr{\pa}{\pa \xi}-\fr{\pa}{\pa \xi^{\prime}}
\right)^m \left(\fr{\pa}{\pa \eta}-\fr{\pa}{\pa \eta^{\prime}}\right)^n
\left(\fr{\pa}{\pa \tau}-\fr{\pa}{\pa \tau^{\prime}}\right)^p
G(\xi,\eta,\tau)F(\xi^{\prime},\eta^{\prime},\tau^{\prime})|_{\xi^{
\prime}=\xi,\eta^{\prime}=\eta,\tau^{\prime}=\tau}.
\eea
%Comparing (55) with (57) we see that there exist the following
%compatible\,(integrable) conditions
%\bea
%& & \ga_{12}=4\beta_1 \beta_2,\\
%& & \fr{\al_{11}}{\ga_{11}-2\beta_1^2}=\fr{\al_{22}}{\ga_{22}-2\beta_2^2}
%    =-\fr{1}{\chi}.
%\eea
In order to get a one-soliton solution we assume
\be
\label{sol1}
F=1+L \exp (\Phi+\Phi^{*}),\,\,\,\,\,\, G=\exp (\Phi)
\ee
with
\be
\label{sol2}
\Phi=(p_R+ip_I)\xi+(q_R+iq_I)\eta+(s_R+is_I)\tau+\Phi_{0R}+i\Phi_{0I},
\ee
where $L,  p_R, p_I, q_R, q_I, s_R, s_I, \Phi_{0R}$ and $\Phi_{0I}$ are
real, yet to be determined constants. Substituting (\ref{sol1}) into
(\ref{form1})-(\ref{form3}) we
obtain the set of algebraic equations
\bea
& & \label{solu1} 8L(\al_{11}p_R^2+\al_{22}q_R^2)-1=0,\\
& & \label{solu2} \ga_{11}(p_R^2-p_I^2)+\ga_{12}(p_R q_R-p_I q_I)
    +\ga_{22}(q_R^2-q_I^2)
    -s_I=0,\\
& & \label{solu3} 2(\ga_{11}p_R p_I+\ga_{22}q_R q_I)
    +\ga_{12}(p_R q_I+p_I q_R)+s_R=0,\\
& & \label{solu4} \chi+8L[(\ga_{11}-2\beta_1^2)p_R^2+(\ga_{22}
    -2\beta_2^2)q_R^2+
    (\ga_{12}-4\beta_1\beta_2)p_R q_R]=0.
\eea
From Eq.(\ref{solu1}) we get
\be
L=\fr{1}{8(\al_{11}p_R^2+\al_{22}q_R^2)}.
\ee
Eqs.(\ref{solu2}) and (\ref{solu3}) give rise to the ``dispersion relations''
\bea
& & s_I=\ga_{11}(p_R^2-p_I^2)+\ga_{12}(p_R q_R-p_I q_I)
        +\ga_{22}(q_R^2-q_I^2),\\
& & s_R=-2(\ga_{11}p_R p_I+\ga_{22}q_R q_I)-\ga_{12}(p_R q_I+p_I q_R),
\eea
with $p_R, p_I, q_R$ and $q_I$ being arbitrary constants. Eqs.(\ref{solu4})
gives a condition for the one-soliton solution.

From (\ref{bili}) and the results given above we have
\bea
& & \label{sing1} A_0=-4(\beta_1p_R+\beta_2q_R)[1+\tanh (\th-\delta_0)],\\
& & \label{sing2} A_1=[2(\al_{11}p_R^2+\al_{22}q_R^2)]^{1/2}{\rm sech}
      (\th-\delta_0)\exp (i\varphi),
\eea
with
$\th=p_R\xi+q_R\eta+s_R\tau+\Phi_{0R}$,
$\varphi=p_I\xi+q_I\eta+s_I\tau+\Phi_{0I}$ and
$\delta_0=(1/2)\log [8(\al_{11}p_R^2+\al_{22}q_R^2)]$\,($\Phi_{0R}$ and
$\Phi_{0I}$ are arbitrary constants). Thus the single-soliton solution
obtained is a line soliton, which consists of two parts, a vibrating wave
packet\,($A_1$, a envelope soliton) and a mean displacement field\,($A_0$,
a kink).

The two-soliton solutions of the Eqs.(\ref{DS11}) and (\ref{DS12}) can be
obtained by chosing
\bea
& & F=1+L_1\exp (\Phi_1+\Phi_1^*)+L_2\exp (\Phi_2+\Phi_2^*)\nonumber \\
& & \hspace{7mm}\label{two1}
    +(L_3+iL_4)\exp (\Phi_1+\Phi_2^*)+(L_3-iL_4)\exp (\Phi_1^*+\Phi_2)
    +L_5\exp (\Phi_1+\Phi_2+\Phi_1^*+\Phi_2^*),\\
& & \label{two2} G=\exp (\Phi_1)+\exp (\Phi_2)
      +(M_1+iM_2)\exp (\Phi_1+\Phi_2+\Phi_1^*)
      +(M_3+iM_4)\exp (\Phi_1+\Phi_2+\Phi_2^*),
\eea
with $\Phi_j=(p_{jR}+ip_{jI})\xi+(q_{jR}+iq_{jI})\eta+(s_{jR}+is_{jI})\tau+
\Phi^0_{jR}+i\Phi^0_{jI}\,(j=1,\,2)$, where $p_{jR}$, $p_{jI}$, $q_{jR}$,
$q_{jI}$, $s_{jR}$, $s_{jI}$, $\Phi^0_{jR}$ and $\Phi^0_{jI}$
are real constants.  When (\ref{two1}) and (\ref{two2}) are substituted into
the bilinear equations (\ref{form1})-(\ref{form3}) we obtain a set of
nonlinear algebraic equations
for the real coefficients $L_j\,(j=1,\,2,...,\,5)$ and
$M_j\,(j=1,\,2,\,3,\,4)$ appearing in (\ref{two1}) and (\ref{two2}).
Solving these equations one can get the expressions of $L_j$ and $M_j$,
as well as the ``dispersion relations'', $s_{jR,I}=s_{jR,I}(p_{jR},p_{jI},
q_{jR},q_{jI})\,(j=1,2)$, which have been given in the Appendix A.
To guarantee (\ref{two1}) and (\ref{two2}) are two-soliton solution,
the following conditions must be imposed
\bea
\label{firstcond}
& & \ga_{12}=4\beta_1 \beta_2,\\
\label{secondcond}
& & \fr{\al_{11}}{\ga_{11}-2\beta_1^2}=\fr{\al_{22}}{\ga_{22}-2\beta_2^2}
    =-\fr{1}{\chi}
\eea
In addition, for $p_{2R}$ and $q_{2R}$, there is a constraint
\be  \label{cons}
\al_{22}(\al_{11}p_{2R}^2+\al_{22}q_{2R}^2)\chi=
(\al_{11}\beta_2^2+\al_{22}\beta_1^2)p_{2R}^2
+2\al_{22}\beta_2^2q_{2R}^2.
\ee
It is easy to show that the integrable conditions of the standard DS
equations\,(i.\,e. the ones amenable to be solved by the
inverse scattering transform) derived in water wave problem are
the particular case of the conditions (\ref{firstcond})
and (\ref{secondcond})\,(see Appendix B).
This fact implies that the GDS equations (\ref{DS11}) and (\ref{DS12})
may  be integrable under the conditions (\ref{firstcond}) and
(\ref{secondcond}).

We note that different equalities in these conditions, however,
reflect different physical properties.
In particular, (\ref{firstcond}) and the first equality in
(\ref{secondcond}) result in
an equation for the wavevector only (i.e. having the form $f(q_1,q_2)=0$
where $f(q_1,q_2)$
does not depend on the lattice parameters, i.e. on $J_2$) for which
the existence of solitons
is possible. Then first two equalities in (\ref{firstcond}) and
(\ref{secondcond}) allows one
to find the particular values of the nonlinear coefficients.
In other words, the above
conditions specify the set of points in the first Brilloun zone,
and necessary values of the
nonlinear forces. What is important for the next consideration,
that such points in the
Brillouin zone do exist. Indeed, as an example we mention that
the above conditions are
satisfied for all points ${\bf q}=(q_1,0)$ and ${\bf q}=(0,q_2)$.

Eqs. (\ref{two1}) and (\ref{two2}) describe two obliquely
interacting solitons in the ($\xi,\eta$)
space. The interaction results in a phase shift\,(i. e. position shift)
for each soliton.

It is possible to get $N$-soliton solutions of the 2D GDS equations
(\ref{DS11}) and (\ref{DS12}) using their bilinear representation,
Eqs.(\ref{form1})-(\ref{form3}) under the integrable conditions
(\ref{firstcond}) and (\ref{secondcond}). We note that
due to the anisotropy inherent in the lattice system\,(i.\,e.
$\beta_2\ga_{12}\neq 0$), the existence of the two-soliton solution
needs the condition
$\ga_{12}=4\beta_1\beta_2$\,(Eq.(\ref{firstcond})\,), which is absent for isotropic
systems\,(e.\,g. water waves).

%%%%%%%%%%%%%%%%%%%%%%%%%%%%%%%%%%%%%%%%%%%%%%%%%%%%%%%
\section{Conclusion}

Using a quasi-discrete multiple-scale method we have derived   the
envelope equations of weakly nonlinear modulations of
$N$-dimensional lattice waves. The equations are obtained for the
case of interaction of a highly frequency mode with a
long-wave-length acoustic one (also called mean field) and can be
classified as generalized Davey-Stewartson equations.
In the case at hand, due to the
anisotropy of the lattice system, the GDS equations in two dimensions
are reduced
either to the DS equations or to a form which does not appears in
the theory of water waves\cite{dav}. The mean field coupled to
the oscillatory short wave packet results from the cubic
interatomic potential in the lattice.  Additionally, generalized
Kadomtsev-Petviashvili equations describing the evolution of a
long wavelength acoustic mode in the lattice are also presented.
We have also studied the modulation instability of Stokes waves and
provided some exact soliton solutions for the two-dimensional
GDS equations based on Hirota's bilinear transformation method.

The results reported here recover the known ones in
one-dimensional systems, which give rise to standard lattice
solitons. On the other hand the method can also be used to study
the weakly nonlinear modulations of the wave packets in vector
lattices or in lattices with a complex cell. The derivation
procedure involves more cumbersome calculation but the envelope
equations obtained take still a form similar to (\ref{eqa0}) and
(\ref{eqa1}) for high frequency wave packets and (\ref{eq:gkp}) and (\ref{3gkp})
for long wave acoustic modes.

\section{Acknowledgements}

The authors are grateful to Profs. X. B. Hu and S.-Y. Lou
for helping the construction of
the bilinear form of the GDS equations and fruitful discussions on the
soliton solutions.
This work was supported in part by the National Natural Science
Foundation of China, the Trans-Century Training Programme Foundation
for the
Talents of the Education Ministry of China, the grants from the
Hong Kong Research Grants Council(RGC), the Hong Kong Baptist
University Faculty Research Grant(FRG), by the FEDER and
Program PRAXIS XXI, No Praxis/P/Fis/10279/1998, and by the Programme
"Human Potential - Research Training Networks", contract
No. HPRN-CT-2000-00158..

\newpage
\centerline{\large \bf Appendix A}

The expressions of $L_j$ and $M_j$ for two-soliton solutions
appearing in (\ref{two1}) and (\ref{two2}) are given by
\begin{eqnarray*}
& & L_1=\fr{1}{8(\al_{11}p_{1R}^2+\al_{22}q_{1R}^2)},\\
& & L_2=\fr{1}{8(\al_{11}p_{2R}^2+\al_{22}q_{2R}^2)},\\
& & L_3=-\fr{1}{2} \fr{\al_{11}\Gamma_{-+}^{-}+\al_{22}\Sigma_{-+}^{-}}
    {[\al_{11}\Gamma_{-+}^{-}+\al_{22}\Sigma_{-+}^-]^2
     +4[\al_{11}\Delta_{-+}^p+\al_{22}\Delta_{-+}^q]^2},\\
& & L_4=-\fr{\al_{11}\Delta_{-+}^{p}+\al_{22}\Delta_{-+}^{q}}
    {[\al_{11}\Gamma_{-+}^{-}+\al_{22}\Sigma_{-+}^-]^2
     +4[\al_{11}\Delta_{-+}^p+\al_{22}\Delta_{-+}^q]^2},\\
& & L_5=\fr{1}{64}\fr{A_{5n}}{A_{5d}},\\
& & M_1=\fr{1}{8}\fr{M_{1n}}{M_{1d}},
    \hspace{5mm}M_2=-\fr{1}{2}\fr{M_{2n}}{M_{2d}},
    \hspace{5mm}M_3=\fr{1}{8}\fr{M_{3n}}{M_{3d}},
    \hspace{5mm}M_4=-\fr{1}{2}\fr{M_{4n}}{M_{4d}},\\
& & L_{5n}=\al_{11}^2 (\Gamma_{--}^+)^2+\al_{22}(\Sigma_{--}^+)^2
    +2\al_{11}\al_{22}(\Gamma_{--}^-\Sigma_{--}^-
    +4\Delta_{--}^p \Delta_{--}^q),\\
& & L_{5d}=(\al_{11}p_{1R}^2+\al_{22}q_{1R}^2)
    (\al_{11}p_{2R}^2+\al_{22}q_{2R}^2)\cdot\nonumber \\
& & \hspace{11mm}[\al_{11}^2(\Gamma_{-+}^+)^2+\al_{22}^2(\Sigma_{-+}^+)^2
    +2\al_{11}\al_{22}(\Sigma_{-+}^-\Gamma_{-+}^-
    -4\Delta_{-+}^p\Delta_{-+}^q)],\\
& & M_{1n}=\al_{11}^2\{(p_{1R}^2-p_{2R}^2)^2+(p_{1I}-p_{2I})^2
    [(p_{1I}-p_{2I})^2-2(3p_{1R}^2-p_{2R}^2)]\}\nonumber \\
& &  \hspace{11mm}+\al_{22}^2\{(q_{1R}^2-q_{2R}^2)^2+(q_{1I}-q_{2I})^2
    [(q_{1I}-q_{2I})^2-2(3q_{1R}^2-q_{2R}^2)]\}\nonumber \\
& & \hspace{11mm}+2\al_{11}\al_{22}\{[(p_{1I}-p_{2I})^2-(p_{1R}^2+p_{2R}^2)]
    [(q_{1I}-q_{2I})^2-(q_{1R}^2+q_{2R}^2)] \nonumber \\
& & \hspace{11mm}-4(p_{1I}-p_{2I})(q_{1I}-q_{2I})(p_{1R}q_{1R}-p_{2R}q_{2R})
    -4q_{1R}q_{2R}p_{1R}p_{2R}\},\\
& & M_{1d}=M_{2d}=(\al_{11}p_{1R}^2+\al_{22}q_{1R}^2)
    [\al_{11}^2(\Gamma_{-+}^+)^2+\al_{22}(\Sigma_{-+}^+)^2
    +2\al_{11}\al_{22}(\Gamma_{-+}^-\Sigma_{-+}^-
    +4\Delta_{-+}^p\Delta_{-+}^q)],\\
& & M_{2n}=\al_{11}^2p_{1R}(p_{1I}-p_{2I})[(p_{1I}-p_{2I})^2-p_{1R}^2+p_{2R}^2]
    +\al_{22}q_{1R}(q_{1I}-q_{2I})
    [(q_{1I}-q_{2I})^2-q_{1R}^2+q_{2R}^2]\nonumber \\
& & \hspace{11mm}+\al_{11}\al_{22}\{(q_{1I}-q_{2I})[q_{1R}\Gamma_{-+}^-
    +2p_{1R}p_{2R}(q_{1I}+q_{2R})]+(p_{1I}-p_{2I})[p_{1R}\Sigma_{-+}^-
    +2q_{1R}q_{2R}p_{2R}]\},\\
& & M_{3n}=\al_{11}^2\{(p_{1R}^2-p_{2R}^2)^2+(p_{1I}-p_{2I})^2
    [(p_{1I}-p_{2I})^2-2(3p_{2R}^2-p_{1R}^2)]\}\nonumber \\
& &  \hspace{11mm}+\al_{22}^2\{(q_{1R}^2-q_{2R}^2)^2+(q_{1I}-q_{2I})^2
    [(q_{1I}-q_{2I})^2-2(3q_{2R}^2-q_{1R}^2)]\}\nonumber \\
& & \hspace{11mm}+2\al_{11}\al_{22}\{[(p_{1I}-p_{2I})^2-(p_{1R}^2+p_{2R}^2)]
    [(q_{1I}-q_{2I})^2-(q_{1R}^2+q_{2R}^2)] \nonumber \\
& & \hspace{11mm}-4(p_{1I}-p_{2I})(q_{1I}-q_{2I})(p_{1R}q_{1R}-p_{2R}q_{2R})
    -4q_{1R}q_{2R}p_{1R}p_{2R}\},\\
& & M_{3d}=M_{4d}=(\al_{11}p_{2R}^2+\al_{22}q_{2R}^2)
    [\al_{11}^2(\Gamma_{-+}^+)^2+\al_{22}(\Sigma_{-+}^+)^2
    +2\al_{11}\al_{22}(\Gamma_{-+}^-\Sigma_{-+}^-+4\Delta_{-+}^p
    \Delta_{-+}^q)],\\
& & M_{4n}=-\al_{11}^2(p_{1I}-p_{2I})p_{2R}[(p_{1I}-p_{2I})^2+p_{1R}^2
    -p_{2R}^2]
    -\al_{22}^2(q_{1I}-q_{2I})q_{2R}[(q_{1I}-q_{2I})^2+q_{1R}^2
    -q_{2R}^2]\nonumber \\
& & \hspace{11mm}-\al_{11}\al_{22}\{(q_{1I}-q_{2I})[q_{2R}\Gamma_{-+}^-
    +2p_{1R}p_{2R}(q_{1I}+q_{2R})]+(p_{1I}-p_{2I})[p_{2R}\Sigma_{-+}^-
    +2q_{1R}q_{2R}p_{1R}]\},\\
\end{eqnarray*}
where
\begin{eqnarray*}
& & \Gamma_{\si_1\si_2}^{\pm}=(p_{1I}+\si_1 p_{2I})^2\pm
    (p_{1R}+\si_2 p_{2R})^2,\\
& & \Sigma_{\si_1\si_2}^{\pm}=(q_{1I}+\si_1 q_{2I})^2\pm
    (q_{1R}+\si_2 q_{2R})^2,\\
& & \Delta_{\si_1\si_2}^p=(p_{1I}+\si_1 p_{2I})
    (p_{1R}+\si_2 p_{2R}),\\
& & \Delta_{\si_1\si_2}^q=(q_{1I}+\si_1 q_{2I})
    (q_{1R}+\si_2 q_{2R}),
\end{eqnarray*}
with $\si_j=\pm 1\,(j=1, 2)$.

The ``dispersion relations'' are given by
\begin{eqnarray*}
& & s_{1R}=-4\beta_1\beta_2 (p_{1R}q_{1I}+p_{1I}q_{1R})
    +2\beta_2^2\left( \fr{\al_{11}p_{1I}p_{1R}}{\al_{22}}-q_{1I}q_{1R}\right)
    +2\beta_1^2\left(\fr{\al_{22}q_{1I}q_{1R}}{\al_{11}}-p_{1I}p_{1R}
    \right),\\
& & s_{1I}=\fr{1}{\al_{11}\al_{22}}\{[-\al_{11}^2\beta_2^2(p_{1R}^2-p_{1I}^2)
    -\al_{22}^2\beta_1^2(q_{1R}^2-q_{1I}^2)\nonumber \\
& & \hspace{11mm}+\al_{11}\al_{22}[\beta_1^2(p_{1R}^2-p_{1I}^2)
    +4\beta_1\beta_2(p_{1R}q_{1R}-p_{1I}q_{1I})+\beta_2^2(q_{1R}^2
    -q_{1I}^2)]\},\\
& & s_{2R}=-4\beta_1\beta_2 (p_{2R}q_{2I}+p_{2I}q_{2R})
    +2\beta_2^2\left( \fr{\al_{11}p_{2I}p_{2R}}{\al_{22}}-q_{2I}q_{2R}\right)
    +2\beta_1^2\left(\fr{\al_{22}q_{2I}q_{2R}}{\al_{11}}-p_{2I}p_{2R}
    \right),\\
& & s_{2I}=\fr{1}{\al_{11}\al_{22}}\{[-\al_{11}^2\beta_2^2(p_{2R}^2-p_{2I}^2)
    -\al_{22}^2\beta_1^2(q_{2R}^2-q_{2I}^2)\nonumber \\
& & \hspace{11mm}+\al_{11}\al_{22}[\beta_1^2(p_{2R}^2-p_{2I}^2)
    +4\beta_1\beta_2(p_{2R}q_{2R}-p_{2I}q_{2I})+\beta_2^2(q_{2R}^2
    -q_{2I}^2)]\},
\end{eqnarray*}
where $p_{jR}$, $p_{jI}$, $q_{jR}$ and $q_{jI}\,(j=1, 2)$ are arbitrary
constants.
%%%%%%%%%%%%%%%%%%%%%%%%%%%%%%%%%%%%%%%%%%%%%%%%%
\newpage
\centerline{\large \bf Appendix B}

One type of the standard DS equations which can be solved by the inverse
scattering transform is\,(see p.\,240 in Ref.[12] for the case of $r=-q^*$)
\bea
& & \label{sta1} \fr{\pa^2 \phi}{\pa x^2}-\si^2 \fr{\pa^2 \phi}{\pa y^2}
=-2 \fr{\pa^2}{\pa x^2}\left(|q^2|\right),\nonumber\\
& & \label{sta2} i\fr{\pa q}{\pa t}+\fr{1}{2}\si^2 \fr{\pa^2 q}{\pa x^2}
+\fr{1}{2}\fr{\pa^2 q}{\pa y^2}=q \phi+|q|^2 q\nonumber
\eea
with $\si^2=\pm 1$. Taking the transformation
$x\rightarrow \xi$, $y\rightarrow \eta$,
$t\rightarrow \fr{2}{\si^2} \tau$,
$q\rightarrow \fr{1}{\sqrt{2}} A_1$,
and $\phi\rightarrow -\fr{\si^2}{2}\fr{\pa A_0}{\pa \xi}$,
above equations become
\bea
& & \label{sta3} \si^2\fr{\pa^2 A_0}{\pa \xi^2}
    -\fr{\pa^2 A_0}{\pa \eta^2}=2\fr{\pa}{\pa \xi}\left(|A_1|^2\right),
    \nonumber\\
& & \label{sta4} i\fr{\pa A_1}{\pa \tau}+\fr{\pa^2 A_1}{\pa \xi^2}
    +\si^{-2}\fr{\pa^2 A_1}{\pa \eta^2}=\si^{-2}|A_1|^2 A_1
    -A_1\fr{\pa A_0}{\pa \xi}.\nonumber
\eea
Comparing with Eqs.\,(\ref{DS11}) and (\ref{DS12}), for the last two
equations we have
\bea
& & \al_{11}=\si^2,\,\,\,\,\al_{22}=-1,\,\,\,\,
    \beta_1=-1,\,\,\,\,\beta_2=0,\nonumber \\
& & \gamma_{11}=1,\,\,\,\,\gamma_{12}=0,\,\,\,\,
    \gamma_{22}=\si^{-2},\nonumber \\
& & \chi=\si^{-2},\nonumber
\eea
which satisfy the integrable conditions (\ref{firstcond}) and
(\ref{secondcond}).

%%%%%%%%%%%%%%%%%%%%%%%%%%%%%%%%%%%%%%%%%%%%%%%%%%%%%%%%%%%%%%%%%%%%%%%%%%%%%%%%%%%%%

%\end{thebibliography}
\end{document}